\documentclass[onecollarge,natbib]{svjour2}
\bibpunct{[}{]}{;}{n}{}{,} 
\smartqed  
\usepackage{graphicx}
\journalname{Few-Body Systems}

\usepackage{amssymb}

\def\ba{\begin{eqnarray}}
\def\ea{\end{eqnarray}}

\def\sp{\kern +3pt}

\begin{document}

\title{
A Relativistic Model for the Electromagnetic Structure
of Baryons from the 3rd Resonance Region
\thanks{This work was supported by
Brazilian Ministry of Science,
Technology and Innovation (MCTI-Brazil).}
}


\author{G.~Ramalho}


\institute{International Institute of Physics, Federal 
University of Rio Grande do Norte, \\
Campus Lagoa Nova - Anel Vi\'ario da UFRN, 
Lagoa Nova, Natal-RN, 59070-405, Brazil \\
              \email{gilberto.ramalho@iip.ufrn.br}      
}

\date{Received: date / Accepted: date}

\maketitle

\begin{abstract}
We present some predictions for the 
$\gamma^\ast N \to N^\ast$ transition amplitudes, 
where $N$ is the nucleon, and $N^\ast$ is  a nucleon excitation  
from the third resonance region.
First we estimate the transition amplitudes 
associated with the second radial excitation of the nucleon,
interpreted as the $N(1710)$ state, 
using the covariant spectator quark model.
After that, 
we combine some results from the covariant spectator quark model 
with the framework of the single quark transition model, 
to make predictions for the 
$\gamma^\ast N \to N^\ast$ transition amplitudes, where $N^\ast$ is 
a member of the $SU(6)$-multiplet $[70,1^-]$.
The  results for the 
$\gamma^\ast N \to N(1520)$ and  $\gamma^\ast N \to N(1535)$
transition amplitudes are used as input 
to the calculation 
of the amplitudes $A_{1/2}$, $A_{3/2}$, 
associated with the 
 $\gamma^\ast N \to N(1650)$,  
$\gamma^\ast N \to N(1700)$,
$\gamma^\ast N \to \Delta(1620)$,
and $\gamma^\ast N \to \Delta(1700)$ transitions.
Our estimates are compared with the available data.
In order to facilitate the comparison with 
future experimental data at high $Q^2$,
we derived 
also simple parametrizations for the amplitudes,  
compatible with the expected falloff at high $Q^2$.
\end{abstract}

\section{Introduction}
\label{intro}

One of the challenges in the modern physics 
is the description of the internal structure 
of the baryons and mesons.
The electromagnetic structure 
of the nucleon $N$ and the nucleon resonances $N^\ast$ can be accessed 
through the $\gamma^\ast N \to N^\ast$ reactions,
which depend  of the (photon) 
momentum transfer squared $Q^2$~\cite{NSTAR,Aznauryan12,CLAS,MAID,Tiator04}. 
The data associated with those transitions 
are represented in terms of   
helicity amplitudes and have been  
collected in the recent years at Jefferson Lab, 
with increasing $Q^2$~\cite{NSTAR}.
The new data demands the development of theoretical models 
based on the underlying structure 
of quarks and quark-antiquark states (mesons)~\cite{NSTAR,Aznauryan12}.
Those models may be used to guide future 
experiments as the ones planned for the Jlab--12 GeV upgrade, particularly 
for resonances in the second and third resonance region
[energy $W =1.4$--$1.8$ GeV]
(see Fig.~\ref{figSigmaW})~\cite{NSTAR}.

An example of a model appropriated 
for the study of the electromagnetic structure 
of resonances at large $Q^2$ is the 
covariant spectator quark model~\cite{Nucleon,OctetFF,Omega}. 
In the covariant spectator quark model 
the baryons are described in terms of 
covariant wave functions based on quarks 
with internal structure (constituent quarks).
Following previous studies for 
the nucleon and the first radial excitation of 
the nucleon~\cite{Nucleon,Roper}, 
we use the covariant spectator quark model
to calculate the transition amplitudes 
associated with the second radial excitation 
of the nucleon~\cite{Roper2}.

In a different work, we use the results of 
the covariant spectator quark model for the 
$\gamma^\ast N \to N(1520)$ and $\gamma^\ast N \to N(1535)$
transition amplitudes~\cite{S11,D13}
to estimate the transition amplitudes 
associated with four negative parity 
nucleon resonances from the $SU(6)$-multiplet $[70,1^-]$,
in the third resonance region.
This study is possible due to the combination
with the single quark transition model, 
which 
allows the parametrization of 
the amplitudes $A_{1/2}$, $A_{3/2}$ for 
six resonances from the  $SU(6)$-multiplet $[70,1^-]$,
based on only three coefficients
dependent of $Q^2$~\cite{SQTM}.


\section{Covariant spectator quark model}
\label{secCSQM}

\begin{figure}
\vspace{.6cm}
\centering
\includegraphics[width=8.6cm]{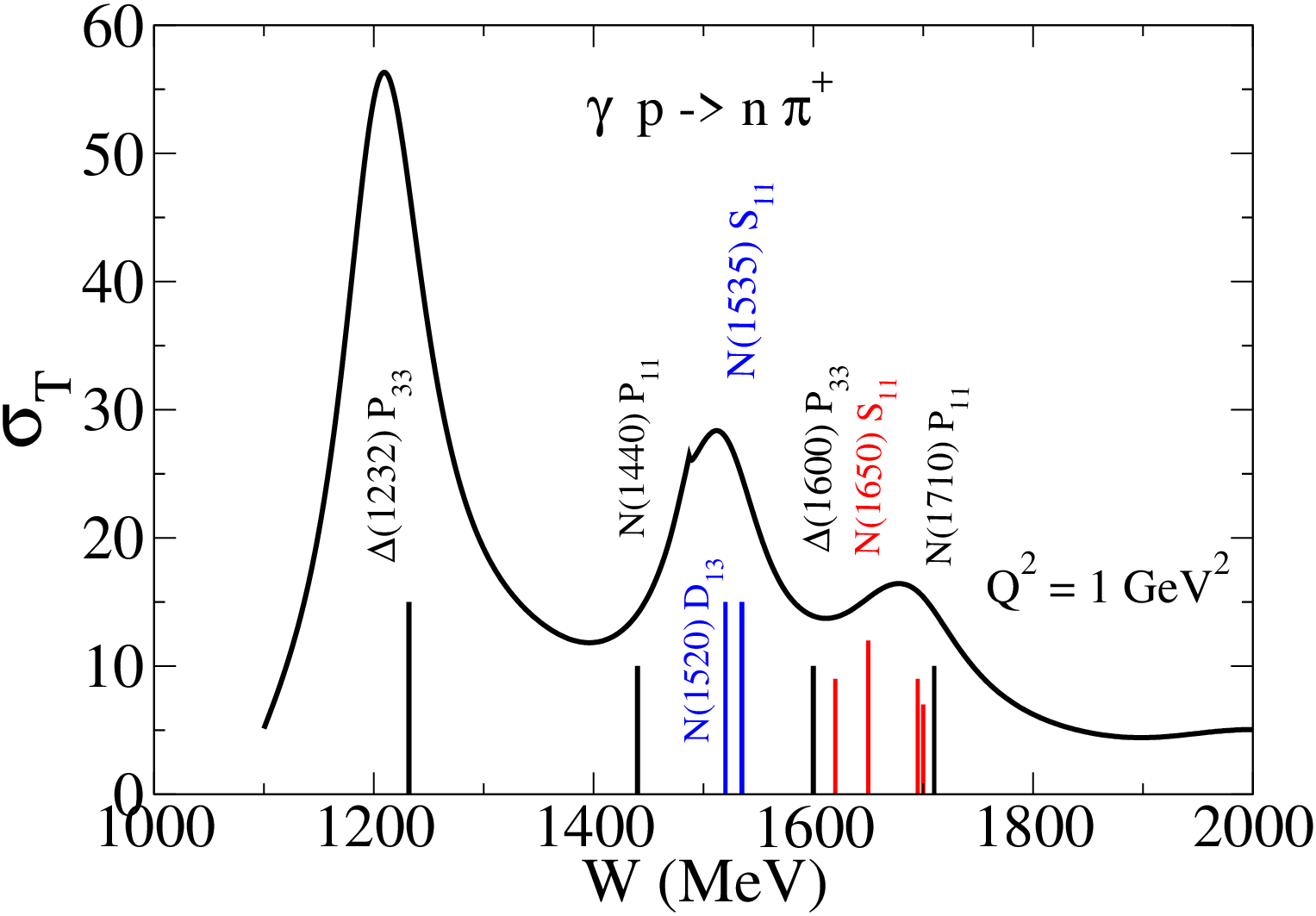} 
\hspace{.6cm}
\begin{minipage}[b]{12pc}
\caption{\label{figSigmaW}
Representation of the $\gamma p \to n \pi^+$ 
cross section.
The graph define the 3 resonance regions.
The vertical lines 
represent resonant states 
described by the covariant spectator quark model, 
including the resonance $N(1710)$.
At red we indicate the states studded 
in this work in the context of the 
single quark transition model.
At blue are the states used as input: $N(1520)$ and $N(1535)$.}
\end{minipage}
\end{figure}

In the covariant spectator quark model,
baryons are treated as
three-quark systems. 
The baryon wave functions are derived from 
the quark states according with the 
$SU(6) \otimes O(3)$ symmetry group.
A quark is off-mass-shell, and  free
to interact with the photon fields,
and other two quarks are on-mass-shell~\cite{Nucleon,OctetFF,Omega}.
Integrating over the quark-pair degrees
of freedom we reduce the baryon to a quark-diquark system,
where the diquark can be represented as
an on-mass-shell spectator particle with an effective
mass $m_D$~\cite{Nucleon,Omega,S11,D13}.

The electromagnetic interaction with the baryons
is described by the photon coupling with
the constituent quarks in the relativistic impulse approximation.
The quark electromagnetic structure is
represented in terms of the quark form
factors parametrized by a vector meson dominance
mechanism~\cite{Nucleon,Omega,Lattice}.
The parametrization of the quark current
was calibrated in the studies of the nucleon form factors data~\cite{Nucleon}
and by the lattice QCD data for the decuplet baryon~\cite{Omega}.
The quark electromagnetic form factors encodes effectively the gluon
and quark-antiquark substructure of the constituent quarks.
The quark current is decomposed 
as $j_q^\mu = j_1 \gamma^\mu  + j_2 \frac{i \sigma^{\mu \nu} q_\nu}{2M}$,
where $j_i$ $(i=1,2)$ are the Dirac and Pauli quark form factors,
and $M$ is the nucleon mass.
In the $SU(2)$-flavor sector
the functions $j_i$ can also be decomposed 
into the isoscalar ($f_{i+}$) and the isovector ($f_{i-}$) 
components: 
$j_i = \frac{1}{6} f_{i+} + \frac{1}{2} f_{i-} \tau_3$,
where $\tau_3$ acts on the isospin states of baryons
(nucleon or resonance).
The details can be found in Ref.~\cite{Nucleon,OctetFF,Omega}.

When the nucleon wave function ($\Psi_N$) 
and the resonance wave function ($\Psi_R$) are both expressed in terms of
the single quark and quark-pair states,
the transition current is calculated
in the relativistic impulse approximation,
integrating over the diquark on-mass-shell momentum,
and summing over the the intermediate diquark 
polarizations~\cite{Nucleon,Omega}.
In the study of inelastic transitions we use 
the Landau prescription to ensure the 
conservation of the transition current~\cite{S11,D13,SQTM}.

Using the relativistic impulse approximation, 
we can express the transition current in terms of the 
quark electromagnetic form factor $f_{i\pm}$ ($i=1,2$)
and the radial wave functions 
$\psi_N$ and $\psi_R$~\cite{Nucleon,S11,D13}.
The radial wave functions are scalar functions that 
depend on the  baryon ($P$) and diquark ($k$) momenta
and parametrize the momentum distributions 
of the quark-diquark systems.
From the transition current we can extract 
the form factors and the helicity transition amplitudes,
defined in the rest frame of the resonance (final state), 
for the reaction under study~\cite{NSTAR,Aznauryan12,S11,D13}.

The covariant spectator quark model was used already 
in the study of several nucleon excitations 
including isospin 1/2 systems 
$N(1410),N(1520),N(1535)$~\cite{Roper,S11,D13}
and the isospin 3/2 systems ~\cite{Lattice,LatticeD,Delta2,Delta1600}.
In Fig.~\ref{figSigmaW}, the position 
of the nucleon excitations are represented and compared 
with the bumps of the cross sections.
The model generalized to the $SU(3)$-flavor sector 
was also used to study the octet and decuplet baryons 
as well as transitions between baryons 
with strange quarks~\cite{Octet2Decuplet,Strange}.
Based on the parametrization of the quark current 
$j_q^\mu$ in terms of the vector meson dominance mechanism, 
the model was  extended to the lattice QCD regime 
(heavy pions and no meson cloud)~\cite{Lattice,LatticeD},
to the nuclear medium~\cite{OctetFF} 
and to the timelike regime~\cite{Timelike}.
The model was also used to study 
the nucleon deep inelastic scattering~\cite{Nucleon,NucleonDIS}
and the axial structure of the octet baryon~\cite{OctetAxial}.

Most of the works refereed below, are based on the 
valence quarks degrees of freedom, 
as consequence of the relativistic impulse approximation.
There are however some processes such as
the meson exchanged between the different quarks
inside the baryon, which cannot be reduced
to simple diagrams with quark dressing.  
Those processes are regarded  
as arising from a meson exchanged between the different quarks inside
the baryon and can be classified as meson cloud corrections 
to the hadronic reactions~\cite{OctetFF,D13,Octet2Decuplet}.
In some cases one can use the covariant spectator quark model 
to infer the effect of the meson cloud effects based 
on empirical information or lattice QCD 
simulations~\cite{S11,Lattice,LatticeD,Octet2Decuplet,OctetAxial,Strange2}.


\section{Resonance $N(1710)$}
\label{secN1710}

We discuss now the  $\gamma^\ast N \to N^\ast$ transitions,
where $N$ is the nucleon state and $N^\ast$  is 
a radial excitation of the nucleon based 
on a $S$-state wave function.
We are excluding where the $J^P = \frac{1}{2}^+$ 
$N^\ast$ states based on spin-isospin wave functions with mixed symmetry.
The $N^\ast$ state share then  
with the nucleon, the structure of spin and isospin,
and differ only in the radial structure 
(radial wave function $\psi_R$).
Therefore, if we exclude
the meson cloud effects, in principle relevant only at low $Q^2$,
we can estimate the transition form factors 
using the formalism already developed for the nucleon~\cite{Nucleon},
replacing the radial wave function of the nucleon,
labeled here as $\psi_{N0}$, in the final state, 
by the radial wave function of the resonances.
Since the spin and isospin structure is the same
for $N$ and $N^\ast$ the orthogonality between the  
radial excitations is a consequence of 
the orthogonality of the radial wave functions.
Labeling the radial wave functions 
of the first and second excitations, as 
$\psi_{N1}$ and $\psi_{N2}$ respectively,
we can express the orthogonality condition 
as~\cite{Roper,Roper2}
\ba
\left.
\int_k \psi_{N1} \psi_{N0} \right|_{Q^2=0}=0,
\hspace{1.2cm}
\left.
\int_k \psi_{N2} \psi_{N0} \right|_{Q^2=0}=0,
\hspace{1.2cm}
\left. 
\int_k \psi_{N2} \psi_{N1} \right|_{Q^2=0}=0,
\label{eqRadial}
\ea
where the subindex $Q^2=0$ indicates that 
the integral is calculated in the limit $Q^2=0$.

The functions $\psi_{N1}, \psi_{N2}$ can be  defined 
in terms of the 
momentum range parameters $\beta_1, \beta_2$
of the nucleon radial wave function, ($\beta_2 > \beta_1$),
where $\beta_1$ regulates the long-range structure
and $\beta_2$ regulates the short-range structure.
If we choose a radial wave function that 
preserves the short-range structure of the 
nucleon wave function, we can determine all parameters 
of the  functions $\psi_{N1}, \psi_{N2}$ using 
Eqs.~(\ref{eqRadial})~\cite{Roper2}.
In this case no parameters have to be adjusted, 
and the model provide true predictions
for the transition form factors or the helicity amplitudes.
This method was already used for the $N(1440)$ state (the Roper),
where we concluded, that, the model gives 
a very good description of the $Q^2 > 1.5$ GeV$^2$ 
data~\cite{Roper,CLAS2,MAID1},
the region where the valence quark degrees of freedom dominate.

\begin{figure}
\vspace{.32cm}
\centering
\includegraphics[width=16pc]{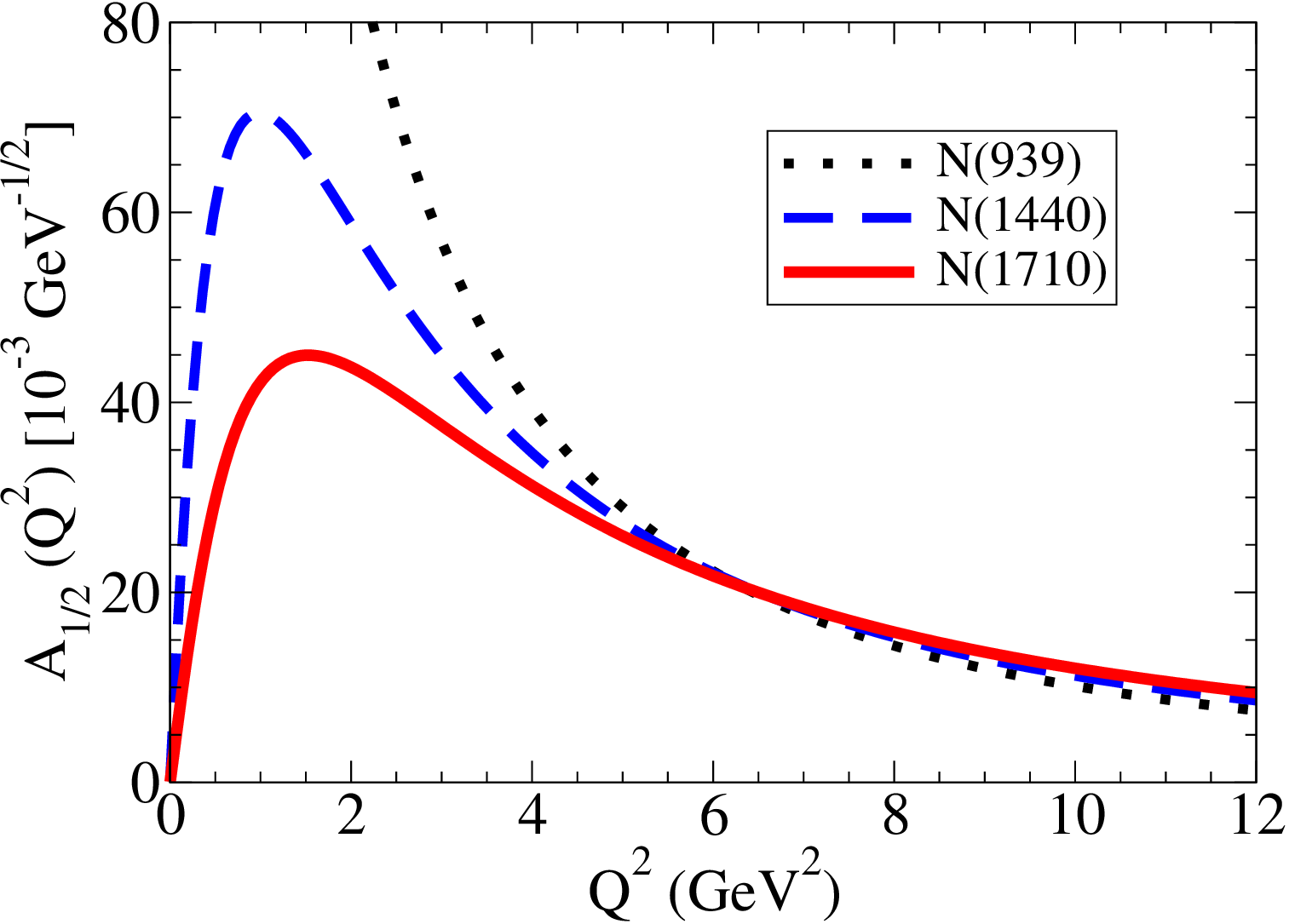} \hspace{1.cm}
\includegraphics[width=16pc]{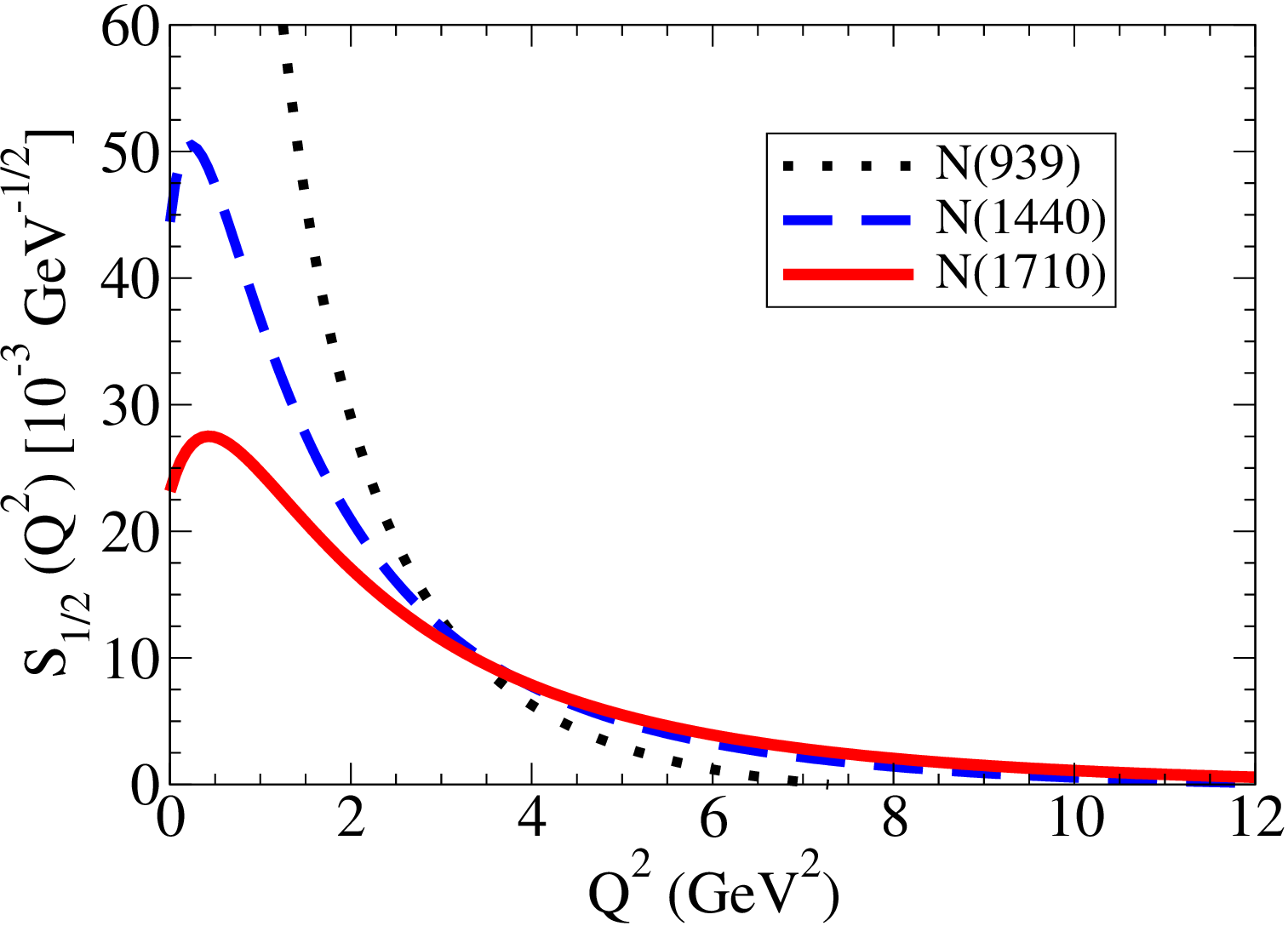}
\caption{Helicity amplitudes for the nucleon, $N(1440)$ and $N(1710)$.
See definition of the nucleon {\it equivalent amplitudes} in the text.}
\label{figN1710-1}
\end{figure}

\begin{figure}[b]
\vspace{.1cm}
\centering
\includegraphics[width=3.0in]{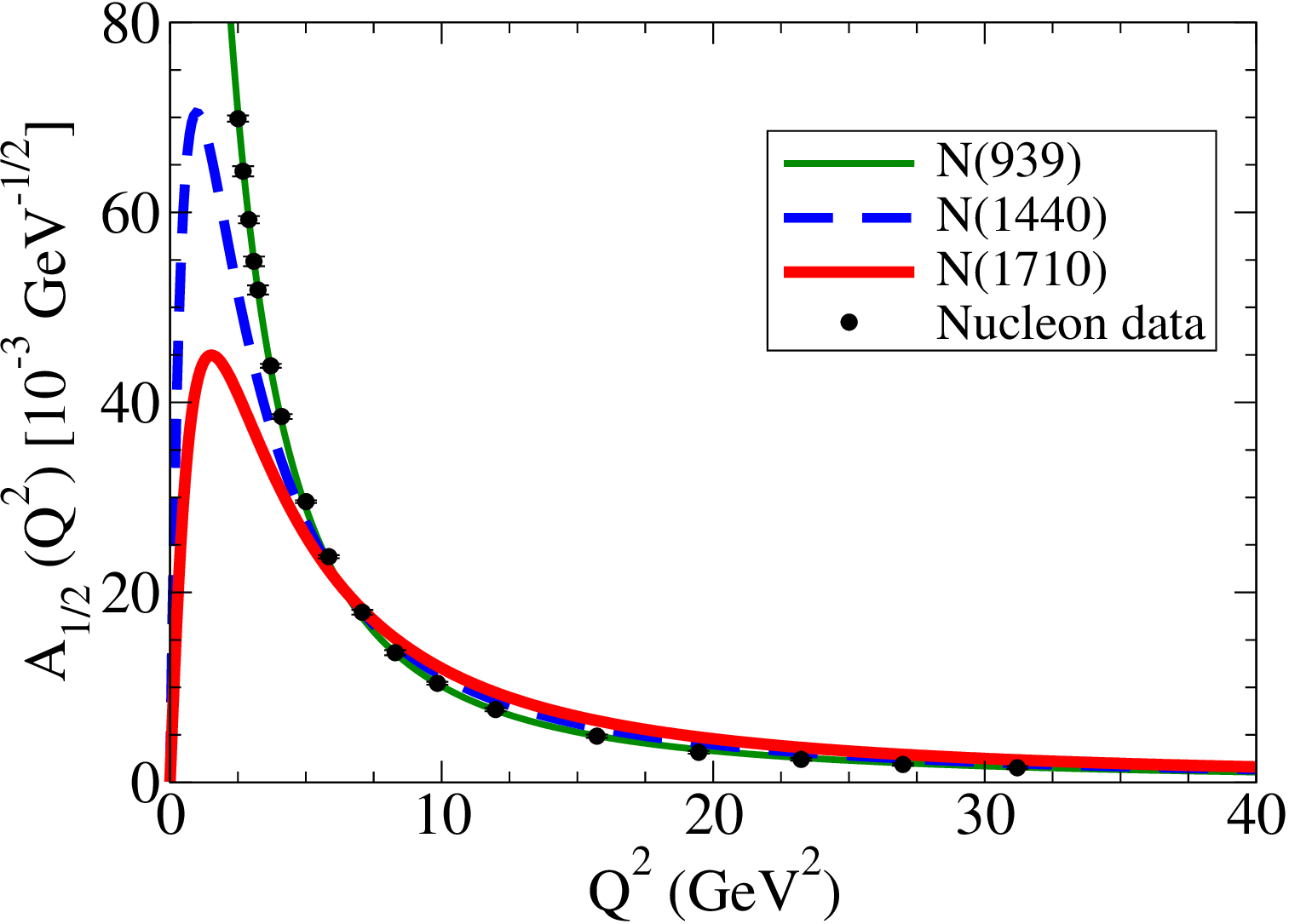}  
\hspace{.3cm}
\begin{minipage}[b]{12pc}\caption{\label{figN1710-2}
Amplitude $A_{1/2}$ for the nucleon, $N(1440)$ 
and $N(1710)$, compared with the data extracted from the nucleon magnetic 
form factor~\cite{Arrington07}.}
\end{minipage}
\end{figure}

We tested then if the model could also be extended for 
the second radial excitation of the nucleon.
Based on the quantum numbers ($P_{11}$),
we assumed that the second radial excitation of the nucleon 
could be the $N(1710)$ state.
The model predictions for the $N(1710)$ transition 
amplitudes are presented in Fig.~\ref{figN1710-1} up to 12 GeV$^2$,
pointing for the upper limit of the Jlab-12 GeV upgrade.
The results are compared with the equivalent results 
for the $N(1440)$ state. 
In order to obtain a comparison between elastic (nucleon) 
and inelastic transition amplitudes
we also present the results of 
the nucleon {\it equivalent amplitudes}. 
The nucleon {\it equivalent amplitudes} are defined by
$A_{1/2} = \sqrt{2} {\cal R} G_M$ 
and $S_{1/2} = \sqrt{2} \frac{{\cal R}}{\sqrt{2}} \sqrt{\frac{1 + \tau}{\tau}} 
G_E$, where $G_M,G_E$ are the nucleon form factors,
$\tau= \frac{Q^2}{4M^2}$, 
and ${\cal R}$ is a function dependent of 
the $N(1440)$ variables (mass $M_R$), 
${\cal R} = \frac{e}{2} \sqrt{\frac{(M_R - M)^2 + Q^2}{M_R M K}}$,
where $K= \frac{M_R^2-M^2}{2 M_R}$ 
 ($e$ is the elementary electric charge). 
In the amplitudes the factor $\sqrt{2}$ was included for convenience.

In Fig.~\ref{figN1710-1}, one can see, that, the results 
for the amplitudes are similar for all the system 
for $Q^2> 4$ GeV$^2$ (large $Q^2$). 
The exception is the result for amplitude $S_{1/2}$ 
for the nucleon, which vanishes for $Q^2 \approx 7$ GeV$^2$,
because $G_E \simeq 0$~\cite{Nucleon}.
One can interpret the approximated convergence of results 
for large $Q^2$ as a consequence of the correlations 
between the excited radial wave functions and the nucleon 
radial wave function~\cite{Roper2}.

Since the {\it equivalent}
amplitude $A_{1/2}$ for the nucleon is known 
already for very large $Q^2$ (using the $G_M$ data), 
this result can be compared with the estimates
of the amplitudes for the $N(1440)$ and $N(1710)$ systems,
for very large $Q^2$, as presented in Fig.~\ref{figN1710-2}.
As shown in the figure, one predicts 
that the amplitude $A_{1/2}$ follows closely 
the {\it equivalent amplitude} of the nucleon, for large $Q^2$.
Future experiments in the range $Q^2=4$--10 GeV$^2$ 
can confirm or deny this prediction.

\begin{figure}
\vspace{.32cm}
\centering
\includegraphics[width=16pc]{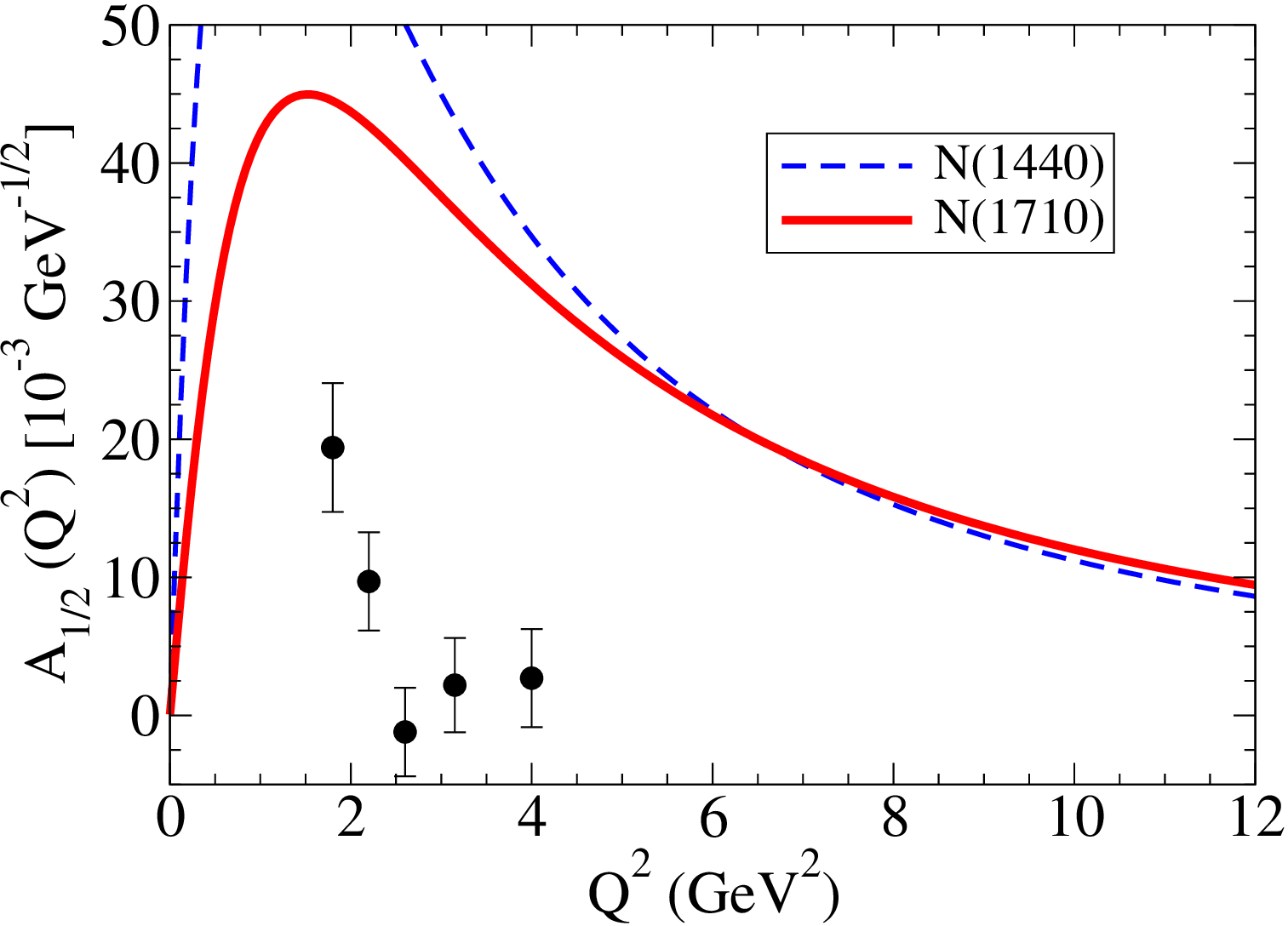}  \hspace{1.cm}
\includegraphics[width=16pc]{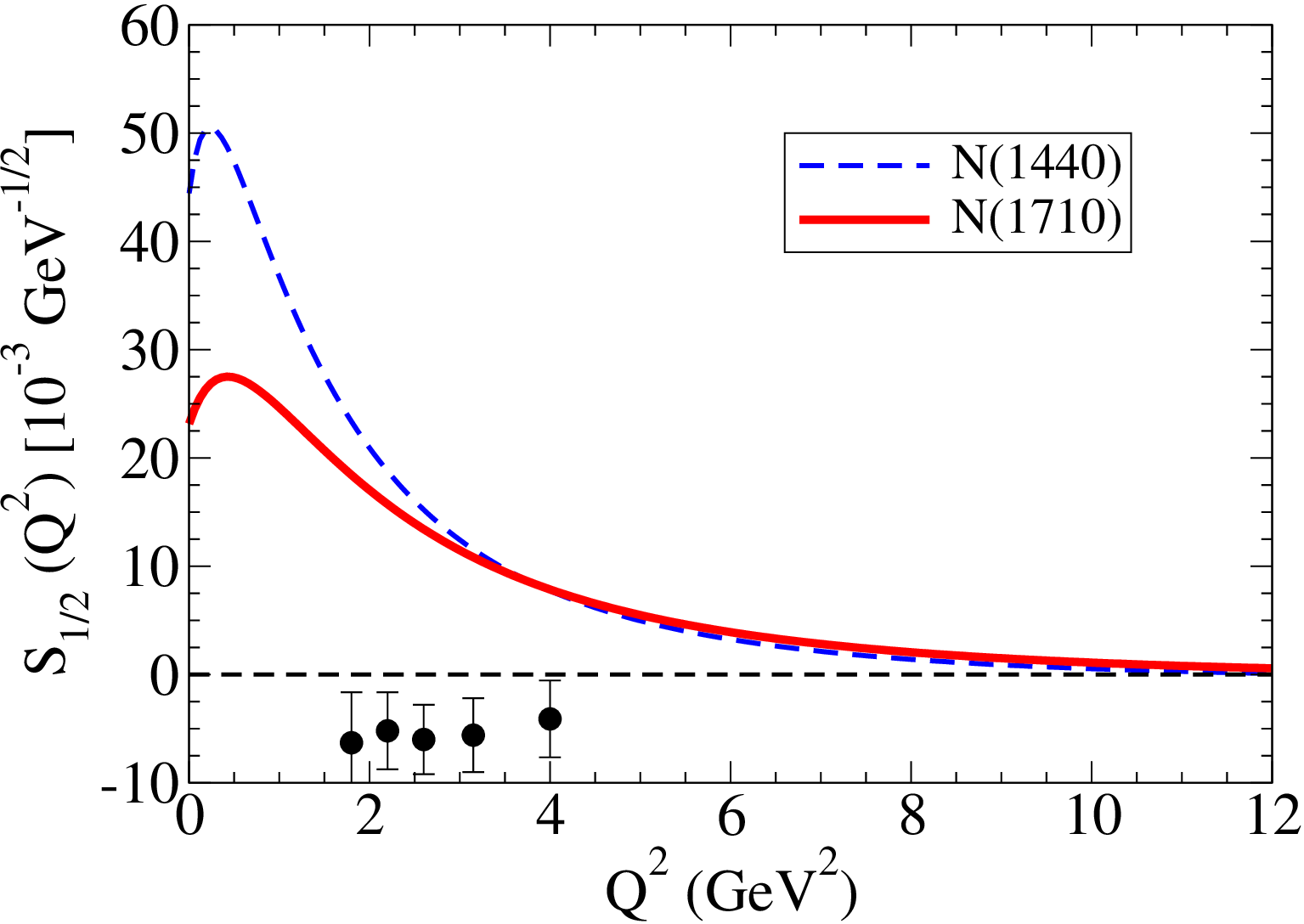}
\caption{$\gamma^\ast N \to N(1710)$ transition amplitudes 
compared with the data from Park {\it et al}~\cite{Park15}.
See discussion in the text.}
\label{figN1710-3}
\end{figure}

The $\gamma^\ast N \to N(1710)$ transition amplitudes 
were determined for the first time in Ref.~\cite{Park15}.
The data is compared with our estimates in Fig.~\ref{figN1710-3}.
From the figure, one can conclude that our estimate 
differs from the data in magnitude for $A_{1/2}$ 
and in sign for $S_{1/2}$, at least for $Q^2 \le 4$ GeV$^2$. 
The new data suggests that, or, our estimate is valid only for 
larger values of $Q^2$, to be confirmed by new data 
for $Q^2 > 4$ GeV$^2$, or, 
our interpretation of $N(1710)$
as the  second radial resonance of the nucleon, is not valid.
It is worth to mention that 
our calculations are consistent with others 
estimates based on the assumption of the second  nucleon radial excitation,
where $A_{1/2} > 0$  and $S_{1/2} > 0$~\cite{Others}.
The signs, and magnitudes of the data, are however consistent 
with the estimates of the hypercentral 
quark model~\cite{Park15,Santopinto12}.

From the theoretical point of view there 
are other possible interpretations 
for the $N(1710)$ state.
Although the interpretation of 
the states from the third resonance region 
based on the quark structure
are partially tentative, 
the state $N(1710)$ can be interpreted 
as an excitation associated with a state with 
mixed symmetry~\cite{PDG,KarlIsgur}.
Several other suggestions have been made
for the composition of the $N(1710)$ state, 
such as the mixture of states $\pi N -\pi \pi N$,
the mixture $\pi N -\sigma N$, a $\sigma_v N$ state,
where $\sigma_v$ represent a vibrational state of the $\sigma$,
and even a $gN$ mixture, where $g$ is a gluon~\cite{Roper2}.
Another possibility is that $N(1710)$ is 
a dynamically generated resonance as suggested by Ref.~\cite{Suzuki10},
although the estimated mass is larger (1820 MeV).
Finally in the interacting quark diquark model 
the $N(1710)$ is described as a quark-axial diquark 
system in a relative $S$-wave configuration.
The prediction of the mass is respectively, 1640 MeV in the 
non relativistic version of the model~\cite{Santopinto05}
and 1776 MeV in the relativistic version of 
the same model~\cite{Santopinto15b}.

Contrarily to the nucleon and the $N(1440)$ 
that are part of the $SU(6)$-multiplet $[56,0^+]$,
the  $N(1710)$ state is more frequently associated 
with the multiplet 
$[70,0^+]$~\cite{Santopinto12,PDG,KarlIsgur,Giannini15}.
In those conditions the second radial excitation 
of the nucleon should correspond to an higher mass resonance.
The state $N(1880)$ listed by the particle data group (PDG)~\cite{PDG} 
is at the moment the best candidate.
Within the quark models it is interesting 
to note that hypercentral quark model 
is the only model which admits 
two radial excitations of the nucleon 
belonging to $[56,0^+]$ 
multiplets~\cite{Santopinto12,Giannini15,Ferraris95}.

We expect, that, future experiments 
provide accurate data for the transverse ($A_{1/2}$) 
and longitudinal ($S_{1/2}$) amplitudes,
which may be used to test the expected 
falloff associated with the valence quark effects:
$A_{1/2} \propto 1/Q^3$  and $S_{1/2} \propto 1/Q^3$, 
for large $Q^2$~\cite{Carlson}.
If this is not case, and $N(1710)$ is in fact 
a mixture of meson-baryon states, 
the $1/Q^3$ falloff cannot be observed, 
and the interpretation of 
$N(1710)$ as a state dominated by  valence quark
has to be questioned.

\vspace{-.1cm}

\section{Resonances from the $[70,1^-]$ $SU(6)$-multiplet}
\label{secSQTM}

The combination of the 
wave functions of a baryon (three-quark system)
given by  $SU(6)\otimes O(3)$ group 
and the description of 
electromagnetic interaction in impulse approximation 
leads to the so-called single quark transition model (SQTM) 
\cite{Burkert03,SQTM-refs}.
In this context {\it single} means 
that only one quark couples with the photon. 
In these conditions the SQTM can be used to 
parametrize the transition current 
between two multiplets,
in an operational form 
that includes only four independent terms,
with coefficients exclusively dependent of $Q^2$.

In particular, the SQTM can be used to parametrize 
the $\gamma^\ast N \to N^\ast$  transitions,
where $N^\ast$ is a $N$ (isospin 1/2) or a $\Delta$ 
(isospin 3/2) state from the
$[70,1^-]$ multiplet, in terms 
on three independent functions of $Q^2$:
$A,B$, and $C$ \cite{Burkert03,SQTM-refs}.
The relations between the functions $A,B$, and $C$
and the amplitudes are presented in the Table~\ref{tableAmp}.
In this analysis we do not take into account 
the constraints at the pseudo-threshold 
when $Q^2 = -(M_R-M)^2$ as in Ref.~\cite{Siegert}.
Using the results for the 
$\gamma^\ast N \to N (1535)$ 
and $\gamma^\ast N \to N(1520)$ amplitudes, 
respectively $A_{1/2}^{S11}$, $A_{1/2}^{D13}$, and $A_{3/2}^{D13}$
in the spectroscopic notation,
we can write, using $\cos \theta_D = 0.995  \simeq 1$ 
\ba
& &
A= 2 \frac{A_{1/2}^{S11}}{\cos \theta_S} 
+ \sqrt{2} A_{1/2}^{D13} +
\sqrt{6} A_{3/2}^{D13}, \hspace{1.5cm} 
B=   2 \frac{A_{1/2}^{S11}}{\cos \theta_S} 
- 2 \sqrt{2} A_{1/2}^{D13}, 
\nonumber \\ 
& &
C= - 2 \frac{A_{1/2}^{S11}}{\cos \theta_S} 
- \sqrt{2} A_{1/2}^{D13} +
\sqrt{6} A_{3/2}^{D13}. 
\label{eqABC}
\ea

We use then the amplitudes $A_{1/2}^{S11}$, $A_{1/2}^{D13}$ 
and  $A_{3/2}^{D13}$, determined by the 
covariant spectator quark model for the $\gamma^\ast N \to N (1520)$
and $\gamma^\ast N \to N(1535)$ transitions~\cite{S11,D13,SQTM}, 
to calculate the coefficients $A,B$ and $C$.
After that we can predict the amplitudes 
associated with the remaining 
transition for  $[70,1^-]$ states, namely 
for the the transitions 
$\gamma^\ast N \to N(1650)$, 
$\gamma^\ast N \to N(1700)$,
$\gamma^\ast N \to \Delta(1620)$ and 
$\gamma^\ast N \to \Delta(1700)$.
Since the covariant spectator quark model 
breaks the $SU(2)$-flavor symmetry, 
we restrict our study to reactions with proton targets
(average on the SQTM coefficients).
Based on the amplitudes used in the calibration 
we expect the estimates to be accurate for $Q^2 \gtrsim 2$ GeV$^2$~\cite{SQTM}.

\begin{table}[t]
\begin{center}
\begin{tabular}{c c c }
\hline
\hline
State & Amplitude &     \\
\hline
$N(1535)$ & $A_{1/2}$ & $\frac{1}{6}(A+B-C) \cos  \theta_S$ \\ [.3cm]
$N(1520)$
& $A_{1/2}$ & $\frac{1}{6\sqrt{2}}(A-2B-C)\cos \theta_D$   \\
& $A_{3/2}$ & $\frac{1}{2\sqrt{6}}(A+ C) \cos \theta_D$   \\ [.3cm]
$N(1650)$ & $A_{1/2}$ & $\frac{1}{6}(A+B-C) \sin  \theta_S$ \\ [.3cm]
$\Delta(1620)$ & $A_{1/2}$ & $\frac{1}{18}(3A-B+C) $ \\ [.3cm]
$N(1700)$ & $A_{1/2}$ &  $\frac{1}{6\sqrt{2}}(A-2B-C)\sin \theta_D$  \\
& $A_{3/2}$ & $\frac{1}{2\sqrt{6}}(A+ C) \sin \theta_D$   \\ [.3cm]
$\Delta(1700)$ & $A_{1/2}$ &  
$\frac{1}{18 \sqrt{2}}(3A+2B+C)$  \\ 
& $A_{3/2}$ & $\frac{1}{6\sqrt{6}}(3A-C) $   \\ [.1cm]
\hline
\hline
\end{tabular}
\hspace{.3cm}
\begin{minipage}[b]{14pc}\caption{\label{tableAmp}
Amplitudes $A_{1/2}$ and $A_{3/2}$ estimated by SQTM 
for the proton targets ($N=p$)~\cite{SQTM,Burkert03}.
The angle $\theta_S$ is the mixing angle associated 
with the 
$N \frac{1}{2}^-$ states ($\theta_S =  31^\circ$).
The angle $\theta_D$ is the mixing angle associated 
with the 
$N \frac{3}{2}^-$ 
states ($\theta_D = 6^\circ$).}
\end{minipage}
\end{center}
\end{table}

From the study of the $\gamma^\ast N \to N (1520)$ 
transition within the covariant spectator quark model 
it is possible to conclude that the contributions 
for the amplitude $A_{3/2}$ due to valence quarks 
are very small~\cite{D13,SQTM}.
This conclusion in consistent 
with others estimates from quark models,
where the valence quark component is about 20--40\%~\cite{D13,Aiello98}.
An accurate description of the  $\gamma^\ast N \to N (1520)$ 
transition requires, then, a significant meson contribution 
for the amplitude $A_{3/2}$, 
as suggested also by the EBAC/Argonne-Osaka model~\cite{EBAC}. 
Therefore, in Ref.~\cite{D13}, we developed an effective 
parametrization of the amplitude $A_{3/2}$, inspired on the meson 
cloud contributions for the $\gamma^\ast N \to \Delta$
transition~\cite{D13,SQTM,Timelike}.

From the relations (\ref{eqABC}), 
we can conclude in the limit where
no meson cloud is considered ($A_{3/2}^{D13}=0$), one has $C=-A$.
The last condition defines the model 1,
dependent only of the parameters $A$ and $B$.
Since, as discussed, the description of the 
 $\gamma^\ast N \to N(1520)$ is limited 
when we neglect the meson cloud contributions 
for $A_{3/2}$, we consider a second model (model 2) 
where in addition to the valence quark contributions 
for the amplitudes $A_{1/2}^{S11}$ and $A_{1/2}^{D13}$, 
we include a parametrization for $A_{3/2}^{D13}$ derived in Ref.~\cite{D13}.

\vspace{.3cm}
\begin{figure}[b]
\vspace{.1cm}
\begin{minipage}{18pc}
\includegraphics[width=16pc]{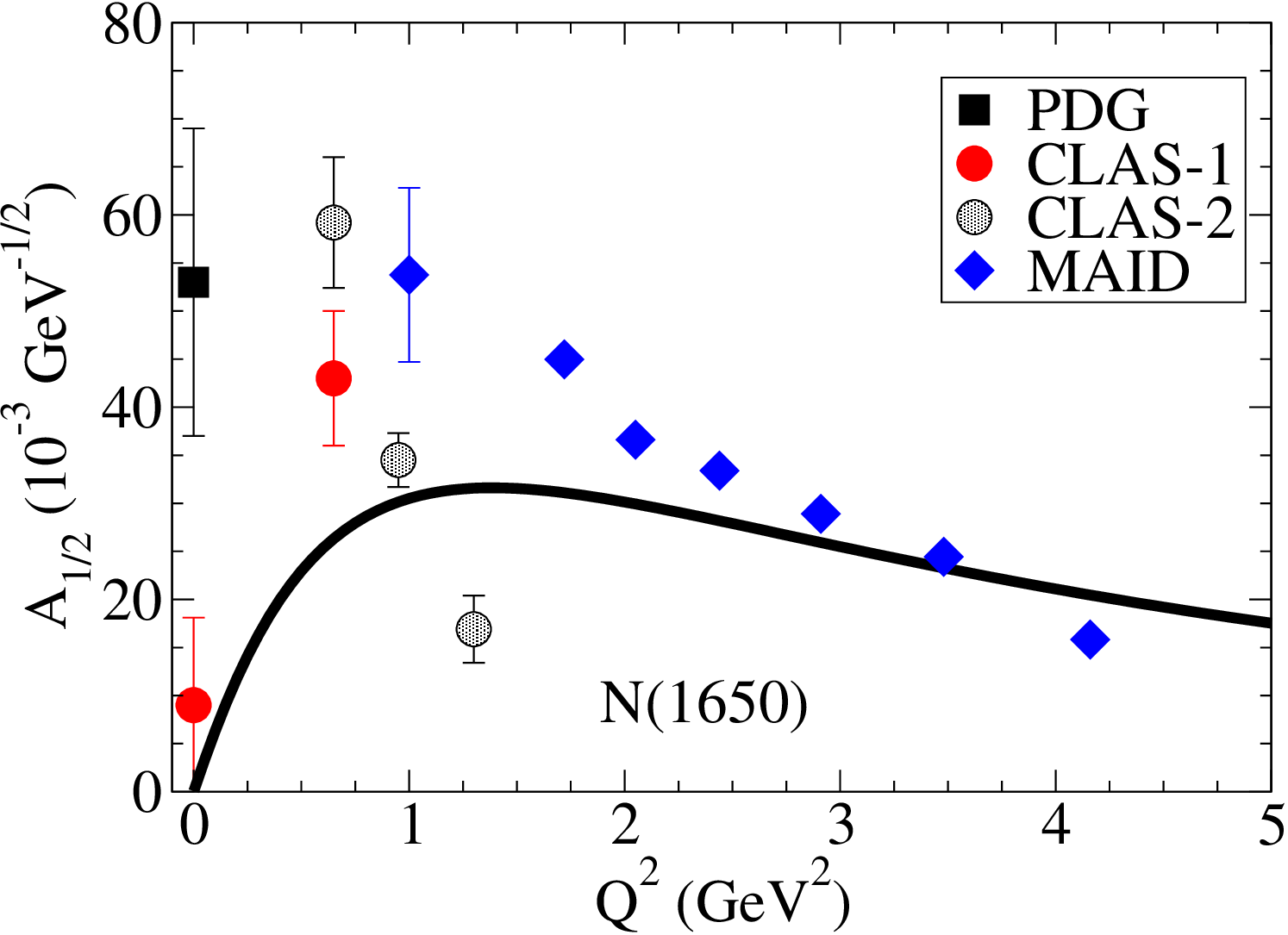}
\caption{\label{figN1650}
Results for $N(1650)$.
The models 1 and 2 gave the same result (solid line).
}
\end{minipage} 
\hspace{2pc}%
\vspace{.6cm}
\begin{minipage}{18pc}
\includegraphics[width=16pc]{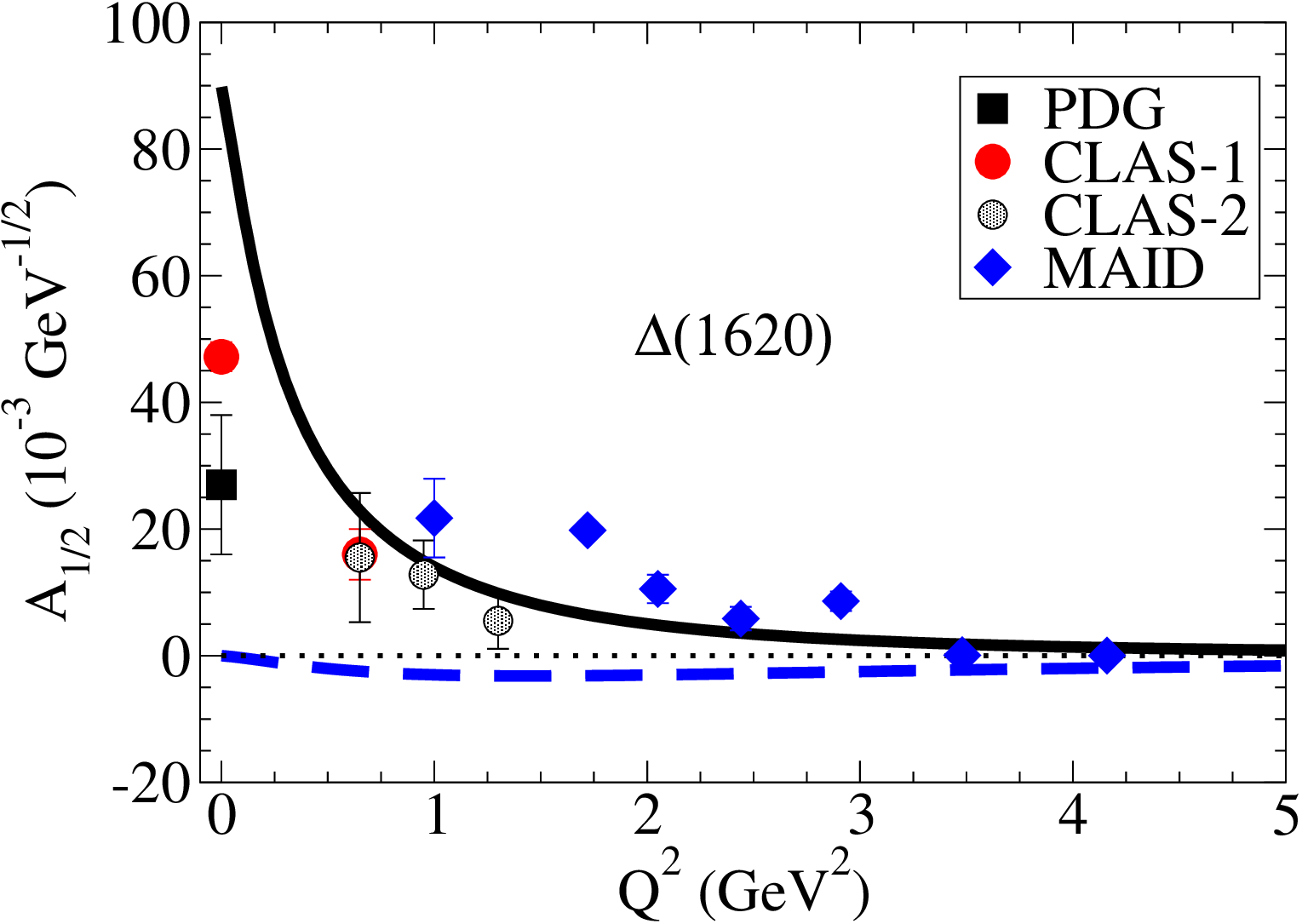}
\caption{\label{figD1620}Results for $\Delta(1620)$.
Model 1 (dashed line) and model 2 (solid-line). }
\end{minipage}
\end{figure}


The models are compared with the available data for 
the amplitudes associated with the $[70,1^-]$ $SU(6)$-multiplet.
The data available for those resonances, 
from CLAS~\cite{CLAS,CLAS2}, MAID~\cite{MAID,MAID1}, 
PDG~\cite{PDG} and others~\cite{Burkert03},
is very scarce particularly for $Q^2 > 1.5$ GeV$^2$.
Nevertheless, we conclude that for the cases
$N(1650)$ and $\Delta(1620)$,
the models 2 gives a good description of the data
(see Figs.~\ref{figN1650} and \ref{figD1620}).

Based on the expected behavior 
for large $Q^2$ given by 
$A_{1/2} \propto 1/Q^3$ 
and $A_{1/2} \propto 1/Q^5$ in accordance
with perturbative QCD arguments~\cite{Carlson},
we parametrize the amplitudes as
\ba
A_{1/2}(Q^2) = D \left(\frac{\Lambda^2}{\Lambda^2 + Q^2}\right)^{3/2},
\hspace{1.5cm}
A_{3/2}(Q^2) = D \left(\frac{\Lambda^2}{\Lambda^2 + Q^2}\right)^{5/2}.
\label{eqLargeQ2}
\ea 
The coefficients $D$ and the cutoff $\Lambda$ 
are determined  in order to be exact for $Q^2 =5$ GeV$^2$.
The results of the parametrizations are in Table~\ref{tableLargeQ2}. 
Those parametrizations may be useful to 
compare with future experiments at large $Q^2$,
as the ones predicted for the Jlab-12 GeV upgrade~\cite{NSTAR}.

For the amplitude $A_{1/2}$ associated 
with the $\Delta(1620)$ state, it is not possible 
to find a parametrization consistent the 
power $3/2$ for the amplitude $A_{1/2}$.
This is because, for that 
particular amplitude, there is a 
partial cancellation between the 
leading terms (on $1/Q^3$) of our $A,B$ and $C$ parametrization, 
due to the difference of sign between 
the amplitudes $A_{1/2}^{S11}$ and $A_{1/2}^{D13}$
used in the determination of the SQTM coefficients 
(see dashed line in Fig.~\ref{figD1620}).
As consequence, the amplitude 
$A_{1/2}$ for the state $\Delta(1620)$ 
is dominated by next leading terms  (on $1/Q^5$)
or contributions due to meson cloud effects ($A_{3/2}^{D13}$).
It is clear in Fig.~\ref{figD1620}, that, 
when we neglect the contributions from $A_{3/2}^{D13}$
the result is almost zero (model 1; dashed line). 
This result shows that in the 
$\gamma^\ast N \to \Delta(1620)$ transition, 
contrarily to what is usually expected, 
there is a strong suppression of the 
valence quark effects for $Q^2=1$--2 GeV$^2$.
A better representation of the 
$\gamma^\ast N \to \Delta(1620)$ data is obtained 
using $A_{1/2} \propto \left( \frac{\Lambda^2}{\Lambda^2 + Q^2}\right)^{5/2}$,
where $\Lambda^2=1$ GeV$^2$ 
(note the power $5/2$, instead of the expected $3/2$).
For a more detailed discussion see~Ref.~\cite{SQTM}.

\begin{table}[t]
\begin{center}
\begin{tabular}{c c c c}
\hline
\hline
State & Amplitude &  $D(10^{-3}$GeV$^{-1/2}$) & $\Lambda^2$(GeV$^2$)   \\ 
\hline 
$N(1650)$ & $A_{1/2}$ & 68.90 &  3.35\\ [.3cm]
$\Delta(1620)$ & $A_{1/2}$ &  ... & \sp ... \\ [.3cm]
$N(1700)$ & $A_{1/2}$ &  $-8.51$\sp\sp & 2.82 \\
& $A_{3/2}$ & 4.36 & 3.61   \\ [.3cm]
$\Delta(1700)$ & $A_{1/2}$ &  39.22  & 2.69 \\ 
& $A_{3/2}$ & 42.15 & 8.42   \\ [.1cm]
\hline
\hline
\end{tabular}
\hspace{.6cm}
\begin{minipage}[b]{14pc}\caption{\label{tableLargeQ2}
Parameters from the high $Q^2$ parametrization 
given by Eqs.~(\ref{eqLargeQ2}).}
\end{minipage}
\end{center}
\end{table}


\section{Summary and conclusions}
\label{secConclusions}

Answering to the challenge raised by recent experimental results 
and by the experiments planed for a near future, 
for resonances in the region $W=$1.6--1.8 GeV, 
with high photon virtualities, 
we present the more recent results 
from the covariant spectator quark model.
The model is covariant, it is 
based on the valence quark degrees of freedom,
and therefore potentially applicable in the region 
of the large energies and momenta.
The estimates for the second radial excitation 
of the nucleon, interpreted as the $N(1710)$ state, 
are still under discussion, both theoretically 
and experimentally. 
At the moment it is more likely 
that the second radial excitation of the nucleon 
correspond to an higher mass resonance.

For the negative parity resonances from 
the $[70,1^-]$ $SU(6)$-multiplet, we provide predictions 
for the transition amplitudes based on the 
combination with the SQTM.
Our predictions compare well with the $Q^2 > 2$ GeV$^2$
data, for $N(1650)$ and $\Delta(1620)$.
As for the remaining resonances, 
the predictions for the transverse amplitudes
$A_{1/2}$ and $A_{3/2}$ have to wait for future high $Q^2$
data, such as the data expected from the Jlab-12 GeV upgrade.

\vspace{-.2cm}




\begin{thebibliography}{20}


\bibitem{NSTAR}
   Aznauryan I.~G., et al.: 
   Studies of Nucleon Resonance Structure in Exclusive Meson Electroproduction.
   Int.\ J.\ Mod.\ Phys.\ E {\bf 22}, 1330015 (2013)



\bibitem{Aznauryan12}
  Aznauryan I.~G.~and Burkert V.~D.: 
  Electroexcitation of nucleon resonances.
  Prog.\ Part.\ Nucl.\ Phys.\  {\bf 67}, 1 (2012)



\bibitem{CLAS}
  Aznauryan I.~G., et al. [CLAS Collaboration]:
  Electroexcitation of nucleon resonances from CLAS data 
  on single pion electroproduction.
  Phys.\ Rev.\  C {\bf 80}, 055203 (2009)



\bibitem{MAID}
  Tiator L., Drechsel D., Kamalov S~S., Vanderhaeghen M.:
  Electromagnetic Excitation of Nucleon Resonances.
  Eur.\ Phys.\ J.\ ST {\bf 198}, 141 (2011)


\bibitem{Tiator04}
   Tiator L., Drechsel D., Kamalov S., Giannini M.~M., Santopinto E.,  
   Vassallo A.:
  Electroproduction of nucleon resonances.
  Eur.\ Phys.\ J.\ A {\bf 19}, 55 (2004)



\bibitem{Nucleon}
  Gross F., Ramalho G., Pe\~na M.~T.:
  A Pure $S$-wave covariant model for the nucleon.
  Phys.\ Rev.\ C {\bf 77}, 015202 (2008);
  Covariant nucleon wave function with $S$, $D$, and $P$-state components.
  Phys.\ Rev.\ D {\bf 85}, 093005 (2012)




\bibitem{OctetFF}
  Ramalho G., Tsushima K.:
  Octet baryon electromagnetic form factors in a relativistic quark model.
  Phys.\ Rev.\ D {\bf 84}, 054014 (2011);
  Ramalho G., Tsushima K., Thomas A.~W.:
  Octet Baryon Electromagnetic form Factors in Nuclear Medium.
  J.\ Phys.\ G {\bf 40}, 015102 (2013)



\bibitem{Omega}
  Ramalho G., Tsushima K., Gross F.:
  A relativistic quark model for the $\Omega^-$ electromagnetic form factors.
  Phys.\ Rev.\  D {\bf 80}, 033004 (2009)





\bibitem{Roper} 
  Ramalho G., Tsushima K.:
  Valence quark contributions for the 
  $\gamma N \to P_{11}(1440)$ form factors.
  Phys.\ Rev.\ D {\bf 81}, 074020 (2010)



\bibitem{Roper2} 
  Ramalho G., Tsushima K.:
  $\gamma^\ast N \to N(1710)$ transition at high momentum transfer.
  Phys.\ Rev.\ D {\bf 89}, 073010 (2014)






  
\bibitem{S11} 
  Ramalho G., Pe\~na M.~T.:
  A covariant model for the $\gamma N \to N(1535)$ 
  transition at high momentum transfer.
  Phys.\ Rev.\ D {\bf 84}, 033007 (2011);
  Ramalho G., Tsushima K.:
  A simple relation between the $\gamma N \to N(1535)$ helicity amplitudes.
  Phys.\ Rev.\ D {\bf 84}, 051301 (2011)


\bibitem{D13}
  Ramalho G., Pe\~na M.~T.:
  $\gamma^\ast N \to N^\ast(1520)$ form factors in the spacelike region.
  Phys.\ Rev.\ D {\bf 89}, 094016 (2014)



\bibitem{SQTM} 
  Ramalho G.: 
  Using the Single Quark Transition Model to predict nucleon 
  resonance amplitudes.
  Phys.\ Rev.\ D {\bf 90}, 033010 (2014)



\bibitem{Lattice} 
  Ramalho G., Pe\~na M.~T.:
  Nucleon and $\gamma N \to \Delta$ lattice form factors 
  in a constituent quark model.
  J.\ Phys.\ G {\bf 36}, 115011 (2009)






\bibitem{LatticeD} 
  Ramalho G., Pe\~na M.~T.:
  Valence quark contribution for the $\gamma N \to \Delta$ 
  quadrupole transition extracted from lattice QCD.
  Phys.\ Rev.\ D {\bf 80}, 013008 (2009)



\bibitem{Delta2} 
  Ramalho G., Pe\~na M.~T., Gross F.:
  A Covariant model for the nucleon and the $\Delta$.
  Eur.\ Phys.\ J.\ A {\bf 36}, 329 (2008);
  $D$-state effects in the electromagnetic $N \Delta$ transition.
  Phys.\ Rev.\ D {\bf 78}, 114017 (2008);
  Electromagnetic form factors of the $\Delta$ with $D$-waves.
  Phys.\ Rev.\ D {\bf 81}, 113011 (2010);
  Ramalho G., Pe\~na M.~T., Stadler A.:
  The shape of the $\Delta$ baryon in a covariant spectator quark model.
  Phys.\ Rev.\ D {\bf 86}, 093022 (2012)


\bibitem{Delta1600} 
  Ramalho G., Tsushima K.:
  A Model for the $\Delta(1600)$ resonance and 
  $\gamma N \to \Delta(1600)$ transition.
  Phys.\ Rev.\ D {\bf 82}, 073007 (2010)







\bibitem{Octet2Decuplet} 
  Ramalho G., Tsushima K.:
  Octet to decuplet electromagnetic transition in a relativistic quark model.
  Phys.\ Rev.\ D {\bf 87}, 093011 (2013);
  %
  What is the role of the meson cloud in the 
  $\Sigma^{*0} \to \gamma \Lambda$ and $\Sigma^\ast \to \gamma \Sigma$ decays?.
  Phys.\ Rev.\ D {\bf 88}, 053002 (2013)




\bibitem{Strange}
   Ramalho G., Tsushima K.:
   Covariant spectator quark model description of the $\gamma^\ast \Lambda \to \Sigma^0$ transition.
   Phys.\ Rev.\ D {\bf 86}, 114030 (2012);
   Ramalho G., Pe\~na M.~T.: 
   Extracting the $\Omega^-$ electric quadrupole moment from lattice QCD data 
   Phys.\ Rev.\ D {\bf 83}, 054011 (2011)





\bibitem{Timelike}
   Ramalho G., Pe\~na M.~T.: 
   Timelike $\gamma^* N \to \Delta$ form factors and Delta Dalitz decay.
   Phys.\ Rev.\ D {\bf 85}, 113014 (2012);
  Ramalho G., Pe\~na M.~T., Weil J., van Hees H., Mosel U.:
  Role of the pion electromagnetic form factor 
  in the $\Delta(1232) \to \gamma^\ast N$ timelike transition.
  Phys.\ Rev.\ D {\bf 93}, 033004 (2016)







\bibitem{NucleonDIS}
  Gross F., Ramalho G., Pe\~na M.~T.:
  Spin and angular momentum in the nucleon.
  Phys.\ Rev.\ D {\bf 85}, 093006 (2012)




\bibitem{OctetAxial}
  Ramalho G., Tsushima K.:
  Axial form factors of the octet baryons in a covariant quark model.
  Phys.~Rev. D {\bf 94}, 014001 (2016)



\bibitem{Strange2}
  Ramalho G., Jido D., Tsushima K.:
  Valence quark and meson cloud contributions for the 
  $\gamma^\ast  \Lambda \to \Lambda^\ast$  and 
  $\gamma^\ast  \Sigma^0 \to \Lambda^\ast$ reactions.
  Phys.\ Rev.\ D {\bf 85}, 093014 (2012)








\bibitem{CLAS2}
   Mokeev V.~I., et al.  [CLAS Collaboration]:
   Experimental Study of the $P_{11}(1440)$ and $D_{13}(1520)$ resonances from CLAS data on $ep \rightarrow e'\pi^{+} \pi^{-} p'$.
   Phys.\ Rev.\ C {\bf 86}, 035203 (2012).
  



\bibitem{MAID1}
  Drechsel D., Kamalov S.~S., Tiator L.: 
  Unitary Isobar Model - MAID2007.
  Eur.\ Phys.\ J.\ A {\bf 34}, 69 (2007);
  Tiator L., Drechsel D., Kamalov S.~S., Vanderhaeghen M.:
  Baryon Resonance Analysis from MAID.
  Chin.\ Phys.\ C {\bf 33}, 1069 (2009);
  Electromagnetic Excitation of Nucleon Resonances.
  Eur.\ Phys.\ J.\ ST {\bf 198}, 141 (2011)





\bibitem{Arrington07} 
  Arrington J., Melnitchouk W., Tjon J.~A.:
  Global analysis of proton elastic form factor data with 
  two-photon exchange corrections.
  Phys.\ Rev.\ C {\bf 76}, 035205 (2007)







\bibitem{Park15} 
  Park K., et al. [CLAS Collaboration]:
  Measurements of $ep \to e^\prime \pi^+n$ at W = 1.6 - 2.0 GeV and extraction of nucleon resonance electrocouplings at CLAS.
  Phys.\ Rev.\ C {\bf 91}, 045203 (2015)



\bibitem{Others}
  Melde T., Plessas W., Sengl B.:
  Quark-Model Identification of Baryon Ground and Resonant States.
  Phys.\ Rev.\ D {\bf 77}, 114002 (2008);
  Ronniger M., Metsch B.~C.:
  Effects of a spin-flavour dependent interaction on light-flavoured 
  baryon helicity amplitudes.
  Eur.\ Phys.\ J.\ A {\bf 49}, 8 (2013)



\bibitem{Santopinto12} 
  Santopinto E., Giannini M.~M.:
  Systematic study of longitudinal and transverse helicity amplitudes in the hypercentral constituent quark model.
  Phys.\ Rev.\ C {\bf 86}, 065202 (2012)






\bibitem{PDG}
  Beringer J., et al. [Particle Data Group Collaboration]:
  Review of Particle Physics (RPP).
  Phys.\ Rev.\ D {\bf 86}, 010001 (2012)




\bibitem{KarlIsgur} 
  Isgur, N., Karl G.:
  P Wave Baryons in the Quark Model.
  Phys.\ Rev.\ D {\bf 18}, 4187 (1978);
  Positive Parity Excited Baryons in a Quark Model with Hyperfine Interactions.
  Phys.\ Rev.\ D {\bf 19}, 2653 (1979);
  Ground State Baryons in a Quark Model with Hyperfine Interactions.
  Phys.\ Rev.\ D {\bf 20}, 1191 (1979)



\bibitem{Suzuki10} 
  Suzuki N., Julia-Diaz B., Kamano H., 
  Lee T.-S.~H., Matsuyama A., Sato T.:
  Disentangling the Dynamical Origin of $P_{11}$ Nucleon Resonances.
  Phys.\ Rev.\ Lett.\  {\bf 104}, 042302 (2010)







\bibitem{Santopinto05} 
  Santopinto E.:
  An Interacting quark-diquark model of baryons.
  Phys.\ Rev.\ C {\bf 72}, 022201 (2005)


\bibitem{Santopinto15b} 
  Santopinto E., Ferretti J.:
  Strange and nonstrange baryon spectra in the relativistic interacting 
  quark-diquark model with a G\"ursey and 
  Radicati-inspired exchange interaction.
  Phys.\ Rev.\ C {\bf 92}, 025202 (2015)




\bibitem{Giannini15} 
  Giannini M.~M., Santopinto E.:
  The hypercentral Constituent Quark Model and its application to baryon properties.
  Chin.\ J.\ Phys.\  {\bf 53}, 020301 (2015)



\bibitem{Ferraris95} 
  Ferraris M., Giannini M.~M., Pizzo M., Santopinto E., Tiator L.:
  A Three body force model for the baryon spectrum.
  Phys.\ Lett.\ B {\bf 364}, 231 (1995)




\bibitem{Carlson}
  Carlson C.~E., Poor J.~L.: 
  Distribution Amplitudes And Electroproduction Of The Delta And Other Low Lying Resonances.
  Phys.\ Rev.\ D {\bf 38}, 2758 (1988)



\bibitem{Burkert03}
  Burkert V.~D., De Vita R., Battaglieri M., 
  Ripani M., Mokeev V.: 
  Single quark transition model analysis of electromagnetic nucleon resonance transitions 
  in the $[70,1^-]$ supermultiplet.
  Phys.\ Rev.\ C {\bf 67}, 035204 (2003)





\bibitem{SQTM-refs}
  Hey A.~J.~G., Weyers J.: 
  Quarks and the helicity structure of photoproduction amplitudes.
  Phys.\ Lett.\ B {\bf 48}, 69 (1974);
  Cottingham W.~N., Dunbar I.~H.: 
  Baryon Multipole Moments In The Single Quark Transition Model.
  Z.\ Phys.\ C {\bf 2}, 41 (1979)



\bibitem{Siegert}
  Ramalho G.: 
  Improved empirical parametrizations of the $\gamma^\ast N \to N(1535)$ 
  transition amplitudes and the  Siegert's theorem.
  Phys.~Lett.~B {\bf 579}, 126 (2016); 
  Improved empirical parametrizations of the $\gamma^\ast N \to \Delta(1232)$ and 
  $\gamma^\ast N \to N(1520)$ transition amplitudes 
  and Siegert's theorem.
  Phys.~Rev.~D {\bf 93}, 113012 (2016); 
  Improved large $N_c$ parametrizations of the $\gamma^\ast N \to \Delta(1232)$ 
  quadrupole form factors and the Siegert's theorem.
  arXiv:1606.03042
 




\bibitem{Aiello98} 
  Aiello A., Giannini M.~M., Santopinto E.:
  Electromagnetic transition form-factors of negative parity nucleon resonances.
  J.\ Phys.\ G {\bf 24}, 753 (1998) 



\bibitem{EBAC}
  Sato T., Lee T.-S.~H.:
  Dynamical Models of the Excitations of Nucleon Resonances.
  J.\ Phys.\ G {\bf 36}, 073001 (2009)


\end{thebibliography}
\end{document}